\documentclass[aip,showpacs,preprintnumbers,amsmath,amssymb,
superscriptaddress,floatfix]{revtex4-1}
\usepackage{movie15}
\usepackage{hyperref}
\usepackage{graphicx}
\usepackage{color}
\usepackage{dcolumn}
\usepackage{bm}
\usepackage[latin1]{inputenc}
\begin{document}
\title{\textcolor{black}{Superconducting wires under simultaneous oscillating
sources: involved magnetic response, dissipation of energy and low pass
filtering}}
\author{H. S. Ruiz}
\email[Electronic address: ]{hsruizr@unizar.es}

\author{A. Bad\'{\i}a\,-\,Maj\'os}
\affiliation{Departamento de F\'{\i}sica de la Materia Condensada 
and Instituto de Ciencia de Materiales de Arag\'on (ICMA),
Universidad de Zaragoza--CSIC,
Mar\'{\i}a de Luna 3, E-50018 Zaragoza, Spain}

\author{Yu. A. Genenko}
\affiliation{Institut f\"{u}r Materialwissenschaft, Technische Universit\"{a}t
Darmstadt, D-64287 Darmstadt, Germany}

\author{S.V. Yampolskii}

\author{H. Rauh}

\date{\today}

\begin{abstract}

\textcolor{black}{Numerical simulations of filamentary type II superconducting wires under
simultaneous AC transport current and oscillating transverse magnetic fields are
performed within the critical state approximation. The time dependences of the
current density profiles, magnetic flux lines, local power dissipation and
magnetic moment are featured. Noticeable non-homogeneous dissipation and field
distortions are displayed. Also, significant differences between the obtained
AC-losses and those predicted by regular approximation formulas are reported.
Finally, an outstanding {\em low pass filtering} effect intrinsic to the
magnetic response of the system is described.}

\end{abstract}
\maketitle

The practical configurations of type-II superconducting wires exhibit complicated
nonlinear and hysteretic behavior under oscillating electromagnetic fields.
Thus, related to the design of power applications with these materials,
considerable efforts have been made to comprehend the factors determining the AC
losses under typical operating conditions. Major features of the macroscopic
electromagnetic behavior have been captured by Bean in the phenomenological
so-called critical state model (CSM).\cite{bean} 
According to this, magnetization currents are induced at
the periphery of the superconductor when external flux variations occur. Such
currents distribute across the section of the sample with a density equal to
so-called critical value at a given temperature and field, $J_{c}$. Although
simple in idealized configurations, the Maxwell equation problem arising from
the CSM statement becomes awkward when realistic systems are considered.

\begin{figure}[b]
\begin{center}
{\includegraphics[width=0.48\textwidth]{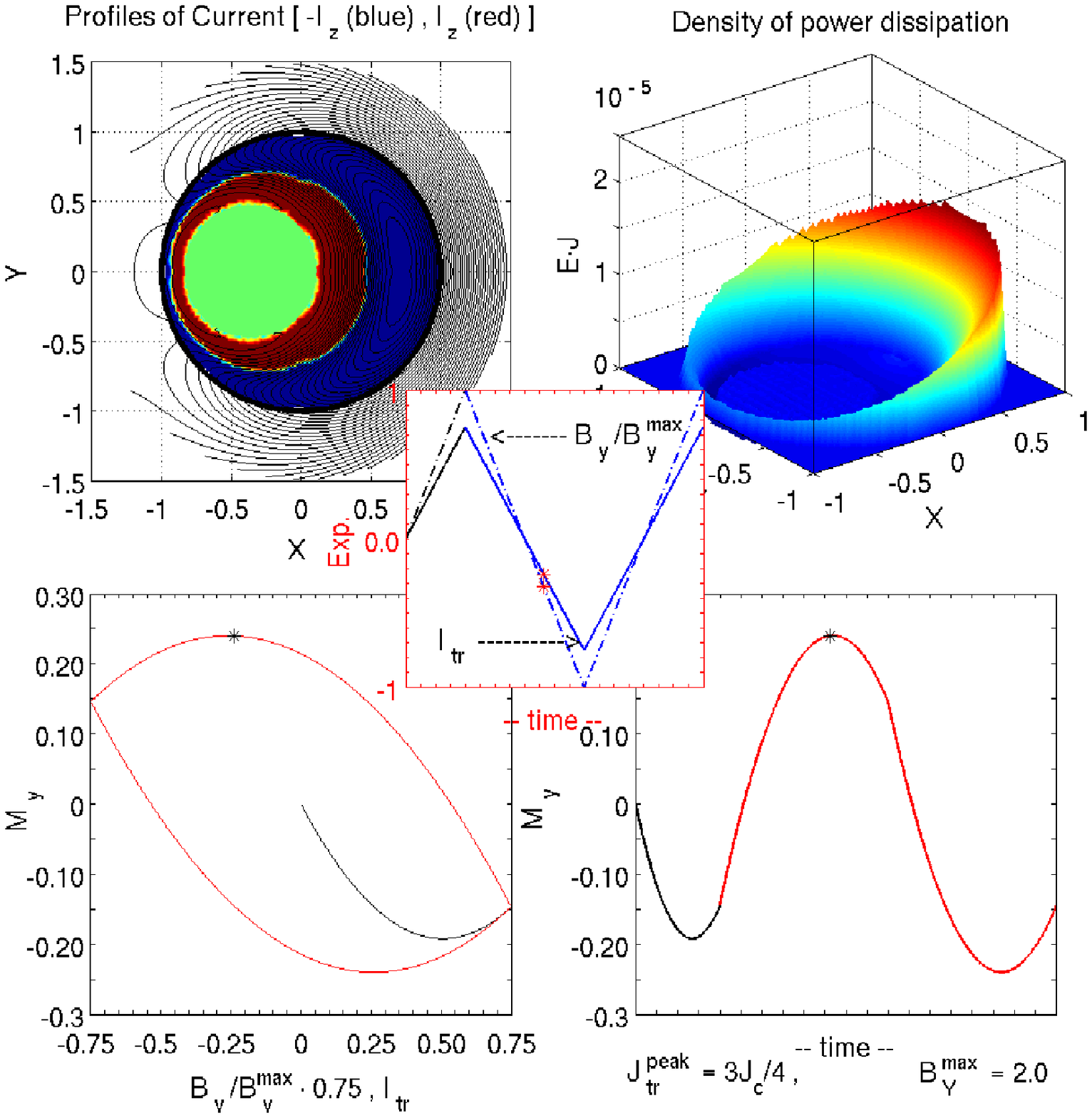}}
\caption{\label{Fig_1}(Color online) Sketch of one of the analized cases. Here,
$B_{y}^{max}=2$, $J_{tr}^{peak}=0.75 J_{c}$, and
the time-step corresponds to $B_{y}=-0.64$, $I_{tr}=-0.24$. Units are
$(\mu_{0}/4\pi)J_{c}R$ for B, $\pi R^{2}J_{c}$ for $I_{tr}$, $J_{c}R^{3}$ for
M, and $(\mu_{0}/4\pi)J_{c}^{2}R^{4}$ for E$\cdotp$J. The experimental process
$(B_{y},I_{tr})$ is summarized in the central plot, and the physics behind
this kind of system can be easily understood analyzing
the \hyperref[Movie]{attached video}.
}
\end{center}
\end{figure}

In this letter, we consider the coupling between simultaneous oscillating
sources acting on a superconducting wire. In particular, the simultaneous action
of an AC transport current and a transverse magnetic field will be studied. 
Notice that such a configuration is a basic model for coil systems in which each
wire is under the action of its neighbors. 
Apparently, the problem is 2D in nature and the main complication is as follows.
When the external sources change in time, a sort of free-boundary problem has to
be solved. Thus, the magnetic flux density variations penetrate from the
sample's surface and an evolutionary flux-front profile is defined. Determining
such front (or the core within) is a main mathematical challenge, and has been
tackled by
several methods.\cite{Ashkin_1979,Ashkin_1986,Carr_1983,Gurevich_1997,Pang_1981,
Telschow_1994, Kuzolev_1995,Gomory_2002,Haken_2002,Karmakar_2003,Rostila_2011} 
As an important handicap, in many cases as the one studied here, the core does
not remain static along the electromagnetic time-varying process, even for the
case of a fully penetrated sample. It is apparent that the widely used concept
of an {\em effective radius} description for the free-boundary can fail markedly
if the core shifts within the sample (as may be observed in Fig.\ref{Fig_1}) and
a {\em center of mass} is not well defined. As a consequence the so-called {\em
front tracking} methods cannot easily guarantee a unique physical solution for
the problem.

Here, we adopt the most popular trend in the analysis of electromagnetic
applications, i.e.: numerical simulations implementing finite-element methods.
As an advantage, such techniques work without explicit inclusion of the free
boundary. The whole superconducting region is involved in the calculation and
the boundary is obtained as a part of the solution. Taking advantage of this
methodology, we have performed a systematic investigation of the electromagnetic
response of the superconducting wire. Some outstanding predictions have been
obtained and are reported here. In particular, (i) we feature the relevance of
localized power dissipation within the sample, (ii) as regards the averaged
physical quantities, we will show that some standard approximations for the
power dissipation per cycle have to be revised, and finally (iii) an intriguing
low pass filtering effect is announced.

Going into detail about our calculations, we use a discrete
formulation that solves Faraday's law iteratively in a mesh of circuits that
carry the macroscopic electric current.\cite{BLR} Calculations are performed
under the material law restriction for the current density, that in this case
reads $|{\bf J}| \leq J_{c}$. Equivalently, one may use {\em conductivity law}
${\bf E}=\rho{\bf J}$ with $\rho({J})=0$ when $|{\bf J}| \leq J_c$ and
$\rho({J})\to\infty$ if $ |{\bf J}| > J_c$.

Under the assumption of translational symmetry, the finite-element
implementation noticeably simplifies. Explicitly, we discretize the sample's
cross-section by a collection of straight infinite elementary wires, each of
them carrying a current $I_{i}=J_{i}S_{i}$ with $J_i$ the current density and
$S_i$ the cross sectional area of the element. As it has been shown in previous
work,\cite{BLR} a variational formulation that suits the discrete modeling is
possible for our electromagnetic problem. Thus, one can show that in
quasi-steady regime (excellent approximation for the large scale application
frequencies) the discrete form of Faraday's law is obtained by minimizing a {\em
magneto-quasi-static} field Lagrangian (of density ${\cal L}=({\bf B}^{t+\delta
t}-{\bf B}^{t})^{2}/2$), coupling successive time layers and under prescribed
sources and material law. In our case, and by using standard electromagnetic
manipulations, the quantity to be minimized becomes
\begin{equation}
\label{eq_minprin}
{\displaystyle \frac{1}{2}}\sum_{i,j}I_{\rm i}M_{ij}I_{\rm j}
-\sum_{i,j}\check{I}_{\rm i}M_{ij}I_{\rm j}
+\mu_{0}\sum_{i}I_{\rm i}S_{i}(A_{e}-\check{A}_{e})\, ,
\end{equation}

with $I_{\rm i}$ the set of unknown currents for the
collection of elements, $M_{ij}$ the mutual inductance matrix between such
elements $i$ and $j$, $\check{I}_{\rm i}$ the solution at the previous time
layer, and $A_{e}$ the vector potential related to external sources (applied
transverse field). Optimization has to be performed under the restriction of
applied transport current in the cross section $\Omega$,
$\sum_{i\in\Omega}I_{\rm i}=I_{tr}$, and for the critical state law $I_{i}\leq
I_{c}$. This has been done by using specialized large scale constrained
minimization algorithms as discussed before.\cite{BLR}

Some technical comments are worth of mention for the case
under study. In general, the electromagnetic manipulations leading to
Eq.(\ref{eq_minprin}) may be done as follows. By using that one can split the
cross section of the cylinder into a high number of elements (wires), it is
justified to assume that the elementary {\em inner} potential is created by a
uniform current density and contributes to the energy as a constant, thus one
can use $M_{ii}=\mu_{0}/4$ in Eq.(\ref{eq_minprin}). On the other hand, using
the logarithmic expression for the two-dimensional Green's function for the
outer potential of the wires one gets
$M_{ij}^{out}=(\mu_{0}/4\pi)\rm{ln}(r_{ij}^{2})$ for the mutual inductances.
These expressions have been obtained under the condition of continuity for  the
magnetic potential but are arbitrary save to a global constant (gauge
invariance). In the absence of transport current, one can show that such
constant may be obviated in the theory. The reason is that, when deriving
Eq.(\ref{eq_minprin}), spatially constant terms from the potentials of the wires
are multiplied by  $\sum_{i}I_{\rm i}$ ($=0$). However, for problems with
transport, unless one cares about such terms, some calculations, as the value of
${\bf E}$ may be tampered. In order to show how this arises, we recall that,
generally speaking, physically acceptable electric fields have to be expressed
in the form $\textbf{E}=-\partial_{t}\textbf{A}-\nabla\phi$, including an {\em
electrostatic like} term. For long wires, it can be argued that $\nabla\phi$ has
to be constant in space, and thus one has
$\textbf{E}=-[\partial_{t}{A}-C(t)]\hat{\bf k}\equiv [\partial_{t}{A'}]\hat{\bf
k}$ where ${A'}$ works as a {\em calibrated} potential.\cite{Prigozhin_2011}
Taking advantage of the fact that arbitrariness is only up to a constant, the
situation may be tackled by progressively {\em determining} $C(t)$ according to
the physical condition $\textbf{E}=0$ at those points where the magnetic flux
does not vary.

Simulations have been performed for the {\em triangular} oscillating process
displayed in Fig.~\ref{Fig_1}. The following quantities have been focused: (i)
magnetic field lines derived from ${\bf B}={\nabla\times}{\bf A}$, (ii) the
sample's magnetic moment (per unit length)
$M=(1/2l)\int_{\Omega}\textbf{r}\times \textbf{J}$, (iii) the local density of
power dissipation ($\textbf{E}\cdot\textbf{J}$), and the hysteretic AC losses
per unit time and volume for cyclic excitations of frequency $\omega$ calculated
as
$L=(\omega/2\pi^{2}R^{2})\oint_{f.c.}dt\int_{\Omega}\textbf{E}\cdot\textbf{J}$.
Here, $f.c.$ denotes a full cycle of the time-varying electromagnetic sources.
Our results, and conclusions are developed along the following paragraphs.

\paragraph{Flux penetration profiles.}

Fig.~\ref{Fig_1} shows some of the results obtained for one of the experimental
processes considered along this work. The dynamics of the
electromagnetic quantities can be followed in detail by means of the attached
video. Several aspects to be noticed are: (i) the position of the flux free
region (core) and the current density profiles are not axially symmetric by
consumption of the magnetization currents. Related to this, (ii) strong
distortions of the magnetic flux lines appear around the superconducting cable.
This is especially notorious when the value of the applied field and the
transport current approach zero. (iii) Outstandingly, the local density of power
dissipated along the process also displays a strong localization. In fact, a
higher heat production (and transfer for constant $T$) is always predicted for
half cross-section of the superconducting wire. We argue that this phenomenon
could increase the quench probability. Furthermore, the proper determination of
the {\em active area} depends on the history of the first branch of the
experimental process, e.g. in Fig.~\ref{Fig_1} a positive slope in both
$B_{y}(t)$ and $I_{tr}(t)$ determines maximal power dissipation towards the
positive x-axis.

\paragraph{Analysis of AC losses.}

\begin{figure*}
\begin{center}
{\includegraphics[width=1.0\textwidth]{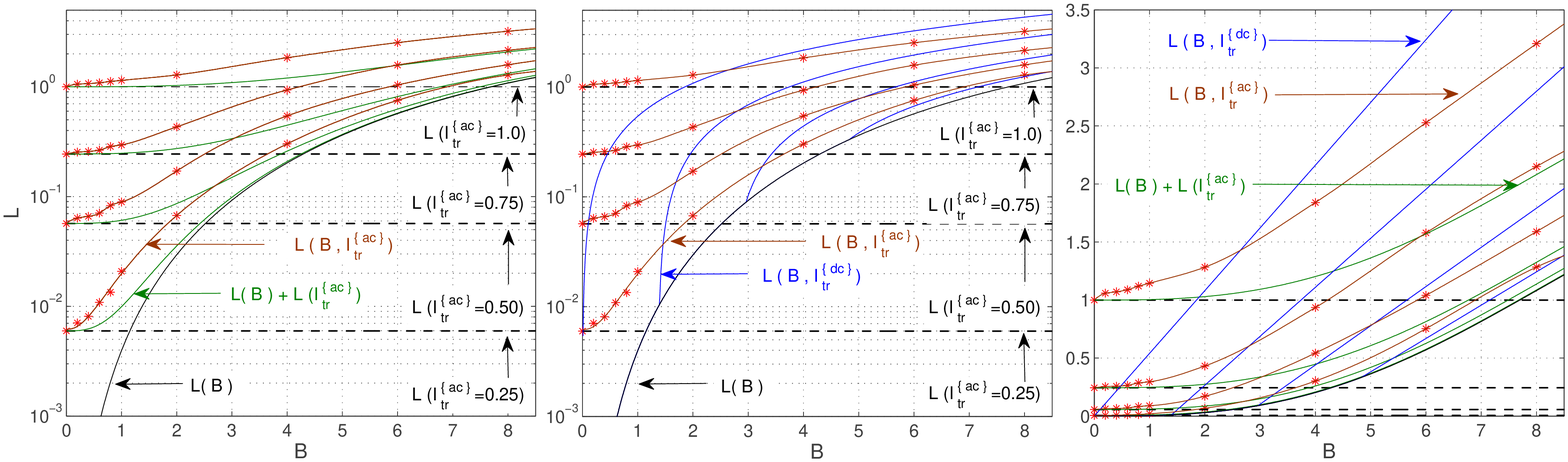}}
\caption{\label{Fig_2}(Color online) Calculated hysteretic ac losses per cycle for simultaneous oscillating transport current
and magnetic flux density, $L(B,I_{tr}^{\{ac\}})$. The different AC transport conditions are defined by the corresponding field amplitudes and labelled peak currents. Our results are directly
compared with the conventional approaches~\cite{Gurevich_1997} for:
(Left) isolated sources $L(B)$ and $L(I_{tr}^{\{ac\}})$, and the
intuitive linear superposition of both of them. (Middle)  oscillating magnetic
flux density under a constant transport current condition 
$L(B,I_{tr}^{\{dc\}})$. (Right) The whole set of results
are also plotted in linear scale. Losses units are
are $(\mu_{0}/4\pi)\omega R^{2} J_{c}^{2}$.}
\end{center}
\end{figure*}

Fig.~\ref{Fig_2} shows the calculated variation of the hysteretic AC
losses for long wires in terms of the amplitude of the coupled electromagnetic sources,
$L(B_{y}^{max},I_{tr}^{peak})$. Our results are compared to those obtained from
several analytical approximations~\cite{Gurevich_1997} customarily applied for
non-coupled periodic sources (see caption at Fig.~\ref{Fig_2}) and {\em
Bean-like superconductors} ($J_{c}=constant$). The main significance of our
results is that linear superposition only makes sense for high magnetic fields
and moderate (or low) currents. This complements previous work on the
rectangular geometry\cite{Pardo_2007} and adds new perspectives on the validity
of approximation formulas. We emphasize the failure of assuming simple linear
superposition of the classical formulas.\cite{Gurevich_1997} In fact, linear
approximations as
$L(B)+L(I_{tr}^{\{ac\}})$ and $L(B,I_{tr}^{\{dc\}})$ (see
Ref.~ \onlinecite{Gurevich_1997}) can notoriously
underestimate or overestimate the real losses. Notice that, even for the best situation,
$L(B)+L(I_{tr}^{\{ac\}})$, at low amplitudes of $B$ the differences between predicted
losses can overcome 100\% (e.g. at  
$L(1,\lesssim0.25))$. Then, they decreases as
the amplitude of $I_{tr}^{\{ac\}}$ increases ($\sim15$\% for
$I_{tr}^{\{ac\}}\simeq1$). On the other hand, if the amplitude of $B$ is as
high as $B_{p}=8$ (the magnetic full penetration field for zero transport current)
the difference can oscillate between $\sim20$\% ($I_{tr}^{\{ac\}}\sim0.25$) and
$\sim55$\% ($I_{tr}^{\{ac\}}\sim1$) or even higher for
$I_{tr}^{\{ac\}}\sim0.75$. 

On average for the extensive set of experiments analyzed, the highest differences $(\sim100\%)$
are observed around the condition $(B_{p}/2,I_{c}/2)$. Specially at low values
of $B$, the differences between analytical approaches and our numerical
calculation can overcome 100\%. In conclusion, for a proper
determination of the hysteretic AC losses in systems with coupled
electromagnetic sources $(B,I_{tr}^{\{ac\}})$ somehow sophisticated analysis
resources are needed, even for relatively simple configurations as the one
studied here.

\paragraph{Magnetic moment cycles (low pass filtering)}

\begin{figure*}
\begin{center}
{\includegraphics[width=0.8\textwidth]{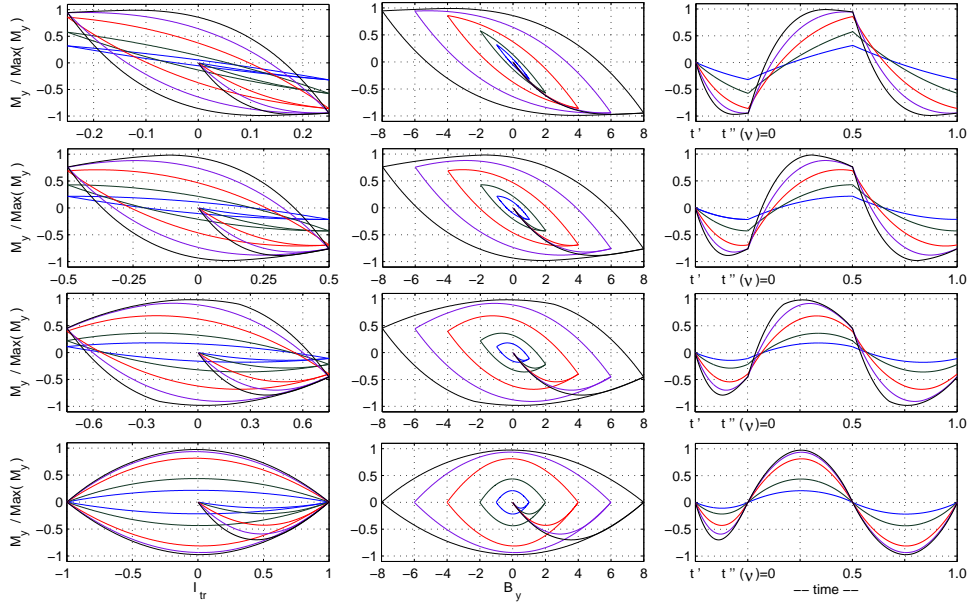}}
\caption{\label{Fig_3}(Color online) The renormalized magnetic moment
$M_{y}/Max(M_{y})$ as a function of the applied sources, $I_{tr}$ (left),
$B_{y}$ (middle), and its temporal evolution (right). Several
experiments are shown according to the maximal intensity of the applied sources
as was sketched in Fig.~\ref{Fig_1}. Indeed, each row corresponds to a singular
choice of $I_{tr}^{peak}=$0.25, 0.5, 0.75, and 1.0. The amplitude of
the oscillating magnetic field has been assumed also to change to consider the
whole spectrum of possible experiments. Here we show the corresponding
results for $B_{y}^{max}=$1, 2, 4, 6, and 8, as can be deduced directly from
the plots.}
\end{center}
\end{figure*}

Fig.~\ref{Fig_3} shows the dynamics of the magnetic moment component along the external field $M_y$ 
in terms of the evolution of the magnetic sources $B_{y}(t)$, $I_{tr}(t)$. Notice that, only for small
values of the applied transport current, nearly standard {\em Bean-like} loops
are obtained. In particular, we recall that the flat saturation behavior of
$M_y$ for high field values progressively disappears by increasing the value
$I_{tr}^{peak}$. Remarkably, this phenomenon ends up with a symmetrization of
the loop (both as a function of $I_{tr}$ and $B_y$), that exhibits a
characteristic eye-shape. As a consequence of such process, an outstanding low
pass filtering effect is predicted for the experimental situation described in
this paper, which may lead to envisage new applications for superconducting
wires. Thus, a plot of the induced magnetic moment as a function of time
(rightmost column in Fig.~\ref{Fig_3}) demonstrates that the triangular input
excitation produces a nearly perfect sinusoidal output $M_{y}(t)$.

\paragraph{Conclusion.}

In this work, we have investigated the electromagnetic response of a
superconducting circular wire under simultaneous AC sources through a numerical
implementation of Bean's critical state model. The main physical assumptions are
an infinitely steep $E(J)$ law that goes from zero to infinity at the critical
point $J_c$. In simple systems, these assumptions are known to lead to
hysteretic (rate independent) losses with simple relations between monotonic and
cyclic quantities. Here, we show that even for the 2D configuration studied, one
can find intriguing phenomena, unexpected in simplified models that are based on
plain linear superposition. Thus, we have obtained notorious localization
effects in the density of power dissipation, strong field distortions, important
failures of the customary approximation formulas for the bulk quantities, and
also predict an outstanding low pass filtering effect in the magnetic response.

Funding of this research is gratefully acknowledged.\cite{Acknowledges}

%
%

\newpage
%
%

\begin{figure}[h!]
{\includegraphics[width=0.48\textwidth]{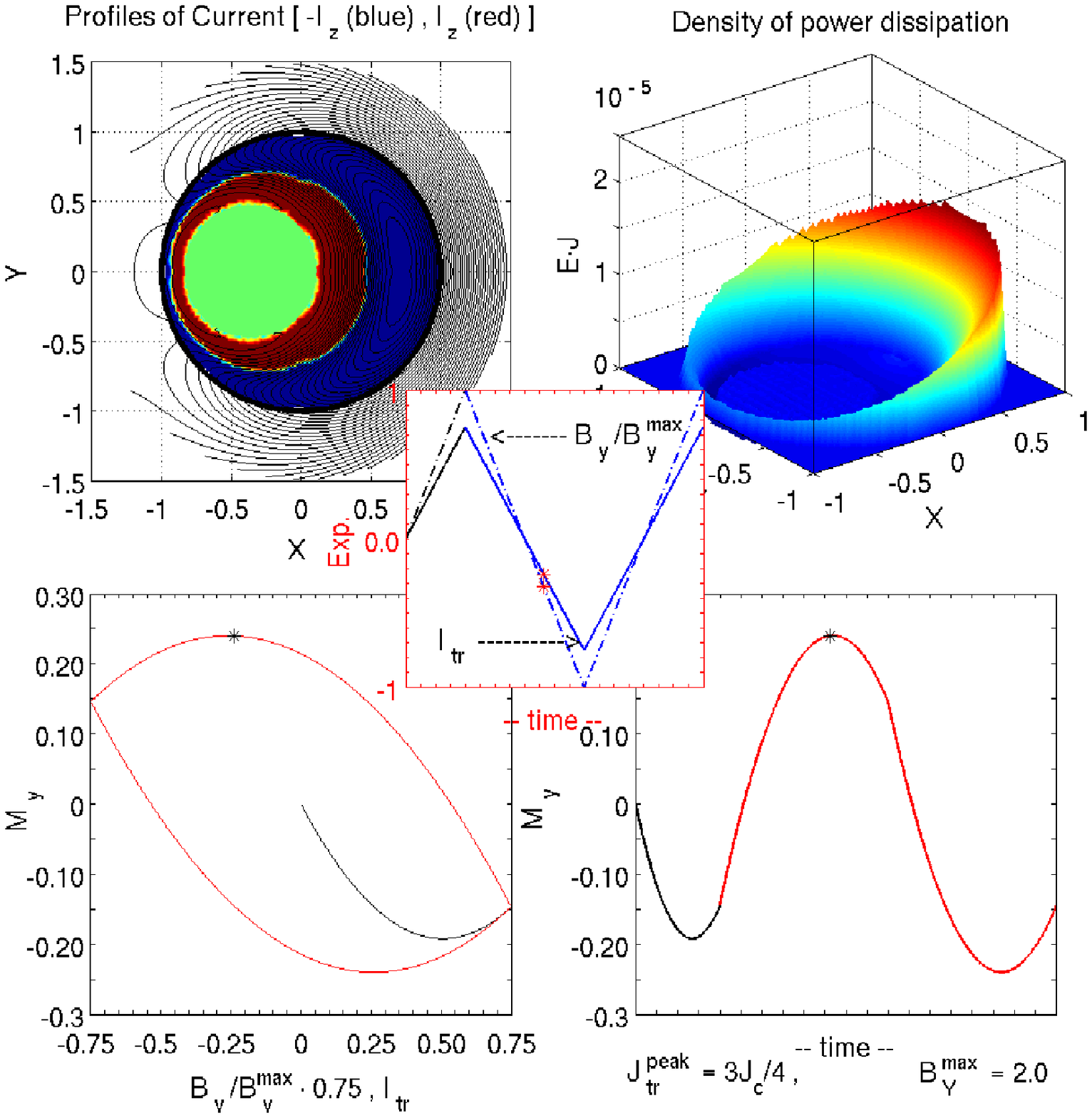}}
\caption{\label{Movie} The video has been uploaded to the Data Conservancy
Pilot Project of arXiv repositories. Also, this and more
videos are available to download in:
\textcolor{blue}{ 
\url{http://www.unizar.es/departamentos/fisica_mat_condensada/people/hsruizr/}
}
\\
\textit{Movie Caption:}
\\
Sketch of one of the analized cases along this
letter. In
this case, $B_{y}^{max}=2$, and $J_{tr}^{peak}=0.75 J_{c}$ (Figure~ \ref{Fig_1}
is a frame of this video). Units are
$(\mu_{0}/4\pi)J_{c}R$ for B, $\pi R^{2}J_{c}$ for $I_{tr}$, $J_{c}R^{3}$ for
M, and $(\mu_{0}/4\pi)J_{c}^{2}R^{4}$ for E$\cdotp$J. Here, we have
assumed a quasistationary time-step defined by the experimental process
$(B_{y},I_{tr})$ which is summarized in the central plot. Top-Left. The magnetic
field lines (projected isolevels of the vector potental over the wire
cross-section) and their corresponding profiles of current are 
shown. The consumption of local magnetization currents and the generated field
distortions around the superconductor wire are directly visualized. 
From time to time, straigth isolines in zones free of current
are ploted as a consequence of the number of isolevels which has been
required to be plotted. They must be understanding only as a visual effect
introduced by the graphical processing. Top-Right: Dynamics of the density of
power dissipation along the cross-section of the superconducting wire. A clear
non symetric distribution of the heat transfer in a cyclic process is always
observed. Bottom-Left: The magnetization loop for the superconducting wire in
terms of the renormalized electromagnetic sources
$(B_{y}/B_{y}^{max}\cdot0.75,0.75)$ is shown. Bottom-Right: The magnetic
moment is shown in terms of the virtual time defined by the dynamics of the
electromagnetic sources. A low pass filtering effect in the magnetic response as
$J_{tr}\rightarrow J_{c}$ is predicted.}
\end{figure}


\end{document}